\documentclass[11pt,onecolumn]{article}
\usepackage{amsmath,amsthm,amssymb,amsfonts,epsfig,graphicx}
\usepackage{color}
\usepackage{dsfont}
\usepackage{url}
\usepackage{array,multirow}
\usepackage{arydshln}
\usepackage{hyperref}

\usepackage{enumitem}
\usepackage[vlined,linesnumbered,ruled,resetcount]{algorithm2e}

\usepackage[left=1.0in,top=1.0in,right=1.05in,bottom=0.5in,nohead,textheight=10in,footskip=0.3in]{geometry}

\newtheorem{theorem}{Theorem}

\newtheorem{lemma}[theorem]{Lemma}

\newtheorem{remark}{Remark}

\SetKwInput{kwMyRepeat}{Repeat}

\begin{document}

\def\myparagraph#1{\vspace{2pt}\noindent\textbf{#1}\enspace\ignorespaces}



\def\DeltaCeil{{\lceil\Delta\rceil}}
\def\TwoDeltaCeil{{\lceil 2\Delta\rceil}}
\def\OnePointFiveDeltaCeil{{\lceil 3\Delta/2\rceil}}

\long\def\ignore#1{}
\def\myps[#1]#2{\includegraphics[#1]{#2}}
\def\etal{{\em et al.}}
\def\Bar#1{{\bar #1}}
\def\br(#1,#2){{\langle #1,#2 \rangle}}
\def\setZ[#1,#2]{{[ #1 .. #2 ]}}
\def\Pr{\mbox{\rm Pr}}
\def\REACHED{\mbox{\tt REACHED}}
\def\AdjustFlow{\mbox{\tt AdjustFlow}}
\def\GetNeighbors{\mbox{\tt GetNeighbors}}
\def\true{\mbox{\tt true}}
\def\false{\mbox{\tt false}}
\def\Process{\mbox{\tt Process}}
\def\ProcessLeft{\mbox{\tt ProcessLeft}}
\def\ProcessRight{\mbox{\tt ProcessRight}}
\def\Add{\mbox{\tt Add}}

\newcommand{\eqdef}{{\stackrel{\mbox{\tiny \tt ~def~}}{=}}}

\def\setof#1{{\left\{#1\right\}}}
\def\suchthat#1#2{\setof{\,#1\mid#2\,}} 
\def\event#1{\setof{#1}}
\def\q={\quad=\quad}
\def\qq={\qquad=\qquad}
\def\calA{{\cal A}}
\def\calB{{\cal B}}
\def\calC{{\cal C}}
\def\calD{{\cal D}}
\def\calE{{\cal E}}
\def\calF{{\cal F}}
\def\calG{{\cal G}}
\def\calI{{\cal I}}
\def\calH{{H}}
\def\calL{{\cal L}}
\def\calN{{\cal N}}
\def\calP{{\cal P}}
\def\calR{{\cal R}}
\def\calS{{\cal S}}
\def\calT{{\cal T}}
\def\calU{{\cal U}}
\def\calV{{\cal V}}
\def\calO{{\cal O}}
\def\calX{{\cal X}}
\def\calY{{\cal Y}}
\def\s{\footnotesize}
\def\calNG{{\cal N_G}}
\def\psfile[#1]#2{}
\def\psfilehere[#1]#2{}
\def\epsfw#1#2{\includegraphics[width=#1\hsize]{#2}}
\def\assign(#1,#2){\langle#1,#2\rangle}
\def\edge(#1,#2){(#1,#2)}
\def\VS{\calV^s}
\def\VT{\calV^t}
\def\slack(#1){\texttt{slack}({#1})}
\def\barslack(#1){\overline{\texttt{slack}}({#1})}
\def\NULL{\texttt{NULL}}
\def\PARENT{\texttt{PARENT}}
\def\GRANDPARENT{\texttt{GRANDPARENT}}
\def\TAIL{\texttt{TAIL}}
\def\HEADORIG{\texttt{HEAD$\_\:$ORIG}}
\def\TAILORIG{\texttt{TAIL$\_\:$ORIG}}
\def\HEAD{\texttt{HEAD}}
\def\CURRENTEDGE{\texttt{CURRENT$\!\_\:$EDGE}}

\def\unitvec(#1){{{\bf u}_{#1}}}
\def\uvec{{\bf u}}
\def\vvec{{\bf v}}
\def\Nvec{{\bf N}}

\newcommand{\bg}{\mbox{$\bf g$}}
\newcommand{\bh}{\mbox{$\bf h$}}

\newcommand{\bx}{\mbox{$x$}}
\newcommand{\by}{\mbox{\boldmath $y$}}
\newcommand{\bz}{\mbox{\boldmath $z$}}
\newcommand{\bu}{\mbox{\boldmath $u$}}
\newcommand{\bv}{\mbox{\boldmath $v$}}
\newcommand{\bw}{\mbox{\boldmath $w$}}
\newcommand{\bvarphi}{\mbox{\boldmath $\varphi$}}

\newcommand\myqed{{}}

\newcommand{\IBFSFS}{{IBFS$^{\mbox{~\!\tiny FS}}$}}
\newcommand{\IBFSAL}{{IBFS$^{\mbox{~\!\tiny AL}}$}}
\newcommand{\BKFS}{{BK$^{\mbox{~\!\tiny FS}}$}}
\newcommand{\BKAL}{{BK$^{\mbox{~\!\tiny AL}}$}}

\newcommand{\XL}{{X_{\le L}}}


\title{\Large\bf  }
\author{}
\title{\Large\bf \vspace{0pt} Multiplicative Weights for Mixed Block Constraints \\ and Maximum Multicommodity Flow}
\author{Vladimir Kolmogorov \\ \normalsize Institute of Science and Technology Austria (ISTA) \\ {\normalsize\tt vnk@ist.ac.at}}
\date{}
\maketitle

\begin{abstract}
We study multiplicative-weights (MWU) algorithms for block-separable feasibility
problems containing both packing and covering constraints.  Each iteration
optimizes a weighted linear objective over one block, either chosen uniformly
at random or visited in a fixed cyclic order.  We show that the cyclic
algorithm returns an $O(\varepsilon)$-approximate solution after 
$O(k\rho\varepsilon^{-2}\log m)$ oracle calls where $k$ is the number of blocks and $\rho$ is the width of the oracle.
The same holds for the randomized version with high probability.
Previously, such bounds were shown when the coupling constraints have the same
sign~\cite{Radzik1997,VillavicencioGrigoriadis1996}.
These analyses critically depended on monotonicity arguments that are not available
in the mixed case.

As an application, we obtain a deterministic
$(1-\varepsilon)$-approximation for maximum multicommodity flow (M-MCF)
using $O(k\varepsilon^{-2}\log m\log(km/\varepsilon))$ approximate
single-commodity min-cost-flow calls.  
 For planar graphs,
and more generally for graph families admitting a near-linear-time
min-cost-flow algorithm, this gives total running time
$\widetilde O(km\varepsilon^{-2})$.  Compared with the
$O(km^{1+o(1)}\varepsilon^{-1})$ algorithm of~\cite{ChenGraurSidford2025},
this improves the dependence on $m$ for these graph families at the
cost of a worse dependence on $\varepsilon$.

Finally, we give an exponentiation-free finite-precision implementation of
the cyclic algorithm.  It represents each internal weight by a
logarithmic-bit mantissa and exponent, while every price supplied to the
block oracle is rational and has a polynomially bounded common denominator.
\end{abstract}

\section{Introduction}
This paper considers the following block-structured feasibility problem:
\begin{equation}\label{eq:block-packing}
 \mbox{find ~~~ }  x=(x_1,\ldots,x_k)\in \calX_1\times\ldots \times \calX_k
 \mbox{ ~~~ satisfying ~~~ } \sum_{i=1}^k A_i x_i\le b
\end{equation}
Here $\calX_i\subset\mathbb R^{n_i}$ is a convex set, $A_i\in\mathbb R^{m\times n_i}$ is a matrix,
and $b\in\{+1,-1\}^m$. 
Thus, we allow both packing constraints of the form $a^\top x\le 1$ and covering constraints of the form $a^\top x\ge 1$.
We are interested in computing {\em $\varepsilon$-approximate solutions}; these are points $x\in \calX_1\times\ldots \times \calX_k$
satisfying $\sum_{i=1}^k A_i x_i\le b+\varepsilon{\bf 1}$.

Let us define $E^+=\{e\in[m]\mid b_e=1\}$ and $E^-=\{e\in[m]\mid b_e=-1\}$.
We assume access to procedure $\mathtt{Oracle}(i,y)$ that has the following specification.
\begin{center}
\setlength{\fboxsep}{8pt}
\fbox{%
\begin{minipage}{0.9\textwidth}
\centering

\raggedright
\textbf{Input.}
A block $i\in[k]$ and a vector $y\in\Delta_m$, where $\Delta_m$ is the
simplex in $\mathbb R^m$.

\medskip
\textbf{Output.}
A point $x_i\in\calX_i$ that satisfies the following:
\begin{subequations}
\begin{align}
y^\top A_i x_i &\le \min_{x\in\calX_i}y^\top A_i x+\varepsilon/k
 \label{eq:approx-oracle}\\
0&\le (A_i x_i)_e \le \rho &&\forall e\in E^+ \label{eq:approx-oracle:b} \\
-\rho&\le (A_i x_i)_e \le 0 &&\forall e\in E^- \label{eq:approx-oracle:c}
\end{align}
\end{subequations}
\end{minipage}%
}
\end{center}
Here $\rho\ge1$ is the width parameter.

We analyze two natural versions of the Multiplicative Weights Update algorithm for solving this problem:
randomized (MWU-R) and deterministic (MWU-D).
The first one selects the block $i\in[k]$ uniformly at random at each step,
while the second processes the blocks in a cyclic, or round-robin, order.
Our main result is 
\begin{theorem}\label{thm:main}
For any value $\varepsilon\in(0,1/6]$, MWU-D either returns an
$O(\varepsilon)$-approximate solution or correctly reports that
\eqref{eq:block-packing} is infeasible after
$O(k\rho\varepsilon^{-2}\ln m)$ oracle calls.  The same holds for MWU-R
after $O(k\rho\varepsilon^{-2}\ln \tfrac {mk}\delta)$ oracle calls with
probability at least $1-\delta$.

Moreover, assuming that all numerical input data and oracle outputs are
rational numbers of polynomial bit length, MWU-D can be implemented so that
every internal weight is represented with
$O(\log(mk\rho/\varepsilon))$ bits.  Its oracle calls can be restricted to vectors
$y=z/q\in\Delta_m$, where $z\in\mathbb Z_{\ge0}^m$, $\sum_ez_e=q$, and
$q\le(mk\rho/\varepsilon)^{O(1)}$.
\end{theorem}

\subsection{Multicommodity flow problem}
To motivate our work, consider two applications: the {\em concurrent
multicommodity flow problem} (C-MCF) and the {\em maximum multicommodity
flow problem} (M-MCF). Given a capacitated graph $G$ and pairs of terminals $\{(s_i,t_i)\}_{i=1}^k$,
the goal is to find $s_i$-$t_i$ flows $x_i$ for each $i\in [k]$ that can be routed in $G$
simultaneously. In their simplest forms, M-MCF asks to maximize $\sum_i {\tt value}(x_i)$
while C-MCF additionally imposes the constraint that ${\tt value}(x_i)$ is the same for all~$i$.

Using a standard reduction to the feasibility problem (with a binary search on the objective value $F$),
both problems can be cast in the form~\eqref{eq:block-packing}. Importantly, we have $b=(1,\ldots,1)$ in the case of C-MCF
and $b=(1,\ldots,1,-1)$ in the case of M-MCF, where the last coordinate
represents the covering constraint $\sum_i{\tt value}(x_i)\ge F$. The oracle for a given commodity $i$ needs to approximately solve a minimum-cost flow problem in a certain graph.

The work of Grigoriadis and Khachiyan~\cite{GrigoriadisKhachiyan1994}
and Leighton et al.~\cite{LeightonEtAl1995} showed that C-MCF can be solved 
by a randomized algorithm using $O(k\varepsilon^{-2}\log (mk))$ oracle calls, either with high probability or in expectation.
As in MWU-R, they select blocks (i.e.\ commodities) randomly.
Radzik then proved in~\cite{Radzik1997} that replacing uniform sampling with a round-robin selection
gives the same bound.
The same cyclic derandomization was extended to general problems of the
form~\eqref{eq:block-packing} with $b=(1,\ldots,1)$ by Villavicencio and
Grigoriadis~\cite{VillavicencioGrigoriadis1996}, who also allowed each
sweep to visit the blocks in an arbitrary order.
Note that the results of~\cite{GrigoriadisKhachiyan1994,LeightonEtAl1995,Radzik1997,VillavicencioGrigoriadis1996,GrigoriadisKhachiyan1996MinCost}
do not cover M-MCF; their analyses rely on the fact
that all components of the right-hand side $b$ in~\eqref{eq:block-packing} have the same signs.

Our main contribution is the analysis of randomized and cyclic one-block-at-a-time
algorithms for mixed packing and covering constraints, with bounds stated in
terms of individual block-oracle calls.
The analyses of
\cite{GrigoriadisKhachiyan1994,LeightonEtAl1995,Radzik1997,VillavicencioGrigoriadis1996,GrigoriadisKhachiyan1996MinCost} obtain this
bound when all coordinates of $b$ have the same sign.  Extending the
randomized argument to mixed signs is relatively simple and requires only an additional one-sided
residual bound. However, the deterministic extension is much more challenging. 
The analyses in~\cite{Radzik1997,VillavicencioGrigoriadis1996}
rely heavily on monotonicity properties that fail in the mixed case.
Our proof uses a very different technique, and is the main technical contribution of this work.

By applying Theorem~\ref{thm:main} to M-MCF and using binary search we obtain the following result.
\begin{theorem}\label{thm:m-mcf}
Consider a directed or undirected M-MCF instance with $m$ edges, $O(m)$ nodes and $k$ commodities
whose capacities and rewards are polynomially-bounded nonnegative integers.
There is a constant $\varepsilon_{\max}>0$ and a deterministic
algorithm that for a given $\varepsilon\in(0,\varepsilon_{\max}]$ returns a feasible flow of value at least
$(1-O(\varepsilon))\mathrm{OPT}$ using
$O(k\varepsilon^{-2}\log m\log(km/\varepsilon))$
single-commodity minimum-cost-flow calls.  Every call has nonnegative
rational costs of bit length $O(\log(mk/\varepsilon))$ and is allowed
to return a feasible flow of cost at most $1+\varepsilon/k$ times the minimum
cost.
\end{theorem}

The current best-known algorithm for $(1-\varepsilon)$-approximate maximum weighted
multicommodity flow, due to Chen, Graur, and Sidford~\cite{ChenGraurSidford2025},
runs in $O(km^{1+o(1)}\varepsilon^{-1})$.
Combining Theorem~\ref{thm:m-mcf} with the
$O(m^{1+o(1)})$-time exact min-cost-flow algorithm of Chen et
al.~\cite{ChenEtAl2025MinCostFlow} instead gives
$\widetilde O(km^{1+o(1)}\varepsilon^{-2})$ time.  
This has a worse dependence on $\varepsilon$ than the algorithm of~\cite{ChenGraurSidford2025},
but leaves a possibility of improving the dependence on $m$ from {\em almost} linear ($O(m^{1+o(1)})$)
to {\em nearly} linear ($\widetilde O(m)$),
if the complexity of an approximate min-cost flow solver is ever improved to nearly linear time.
In fact, a nearly linear time (exact) min-cost flow algorithm is known for planar graphs, or more generally
graph families that admit an efficient vertex separator of size $O(\sqrt{n})$~\cite{DongEtAl2022}.
Consequently, our work yields an $\tilde O(km\varepsilon^{-2})$-time algorithm for $(1-\varepsilon)$-approximate M-MCF in such families.
We remark that our approach is a short black-box reduction to single-commodity min-cost flow, whereas the
better dependence on $\varepsilon$ in the algorithm of~\cite{ChenGraurSidford2025} relies on more specialized machinery.

\subsection{Other related work}

\myparagraph{Multicommodity-flow algorithms}
Existing algorithms for C-MCF and M-MCF can be grouped into three categories: MWU-type algorithms that
use shortest path computations as the underlying oracle~\cite{ShahrokhiMatula1990,KleinPlotkinSteinTardos1994,PlotkinShmoysTardos1995,GargKonemann1998,Fleischer2000,Karakostas2008,Madry2010},
block-MWU algorithms that rely on a min-cost-flow solver~\cite{LeightonEtAl1995,Radzik1997,GrigoriadisKhachiyan1996MinCost},
and  techniques based on optimizing an area-convex functional~\cite{Sherman2017,ChenGraurSidford2025}.
Table~\ref{tab:mcf-history} summarizes the complexities of these approaches.
The general exponential-potential and price-directive decomposition frameworks
underlying the Grigoriadis-Khachiyan flow result are developed
in~\cite{GrigoriadisKhachiyan1995Exponential,GrigoriadisKhachiyan1996Coordination}.

\begin{table}[t]
\centering
\small
\setlength{\tabcolsep}{3pt}
\resizebox{\textwidth}{!}{%
\begin{tabular}{@{}clclll@{}}
\hline
year & work & flags & complexity & restrictions & technique \\
\hline
1987 & Shahrokhi-Matula~\cite{ShahrokhiMatula1990}
  & $\mathrm{C}\phantom{\mathrm{M}}$ & $\operatorname{poly}(n,m,\varepsilon^{-1})$
  & uniform capacities & \multirow{2}{*}{MWU precursor} \\
1990 & Klein et al.~\cite{KleinPlotkinSteinTardos1994}
  & $\mathrm{C}\phantom{\mathrm{M}}$
  & $\widetilde O(km\varepsilon^{-2}+m^2\varepsilon^{-3})$
  & uniform capacities; randomized & \\
\hdashline
1991 & Plotkin-Shmoys-Tardos~\cite{PlotkinShmoysTardos1995}
  & $\hphantom{{}^{*}}\mathrm{CM}^{*}$ & $\widetilde O(km^2\varepsilon^{-2})$
  & randomized & \multirow{6}{*}{MWU} \\
  & & $\hphantom{{}^{*}}\mathrm{CM}^{*}$ & $\widetilde O(k^2m^2\varepsilon^{-2})$
  & & \\
1998 & Garg-K\"onemann~\cite{GargKonemann1998}
  & $\mathrm{CM}$ & $\widetilde O(km^2\varepsilon^{-2})$
  & & \\
1999 & Fleischer~\cite{Fleischer2000}
  & $\phantom{\mathrm{C}}\mathrm{M}$ & $\widetilde O(m^2\varepsilon^{-2})$
  & & \\
2002 & Karakostas~\cite{Karakostas2008}
  & $\mathrm{C}\phantom{\mathrm{M}}$ & $\widetilde O(m^2\varepsilon^{-2})$
  & implicit & \\
2010 & Madry~\cite{Madry2010}
  & $\phantom{\mathrm{C}}\mathrm{M}$ & $\widetilde O(mn\varepsilon^{-2})$
  & randomized & \\
\hline
1991 & Leighton et al.~\cite{LeightonEtAl1995}
  & $\mathrm{C}\phantom{\mathrm{M}}$ & $\widetilde O(k\varepsilon^{-3}T_{\text{mincost}})$
  & randomized & \multirow{6}{*}{block MWU} \\
  & & $\mathrm{C}\phantom{\mathrm{M}}$ &
    $\widetilde O(k^2\varepsilon^{-2}T_{\text{mincost}})$ 
  &  & \\
1995 & Radzik~\cite{Radzik1997}
  & $\mathrm{C}\phantom{\mathrm{M}}$ & $\widetilde O(k\varepsilon^{-2}T_{\text{mincost}})$
  & & \\
1995 & Grigoriadis-Khachiyan~\cite{GrigoriadisKhachiyan1996MinCost}
  & $\hphantom{{}^{*}}\mathrm{C}^{*}\phantom{\mathrm{M}}$ & $\widetilde O(k\varepsilon^{-2}T_{\text{mincost}})$
  & & \\
2026 & {\bf this work} 
  & $\phantom{\mathrm{C}}\mathrm{M}$ & $\widetilde O(k\varepsilon^{-2}T_{\text{mincost}})$
  & & \\
  & & $\phantom{\mathrm{C}}\mathrm{M}$ & $\widetilde O(km\varepsilon^{-2})$
  & efficient vertex separator &  \\
\hline
2017 & Sherman~\cite{Sherman2017}
  & $\mathrm{C}\phantom{\mathrm{M}}$ & $\widetilde O(mk\varepsilon^{-1})$
  & undirected & \multirow{2}{*}{\shortstack{area-convex\\optimization}} \\
2025 & Chen-Graur-Sidford~\cite{ChenGraurSidford2025}
  & $\mathrm{CM}$ & $O(m^{1+o(1)}k\varepsilon^{-1})$
  & & \\
\hline
\end{tabular}
}
\caption{Selected approximation results for concurrent and maximum
multicommodity flow. C/M flags indicate whether the complexity applies to C-MCF, M-MCF, or both.
A superscript $*$ denotes a cost-constrained or minimum-cost variant.  
$T_{\text{mincost}}$ is the complexity of a single-commodity min-cost-flow computation.  The $\widetilde O$ notation also
hides factors logarithmic in $n$.
}
\label{tab:mcf-history}
\end{table}

\myparagraph{Block-oracle models}
Problem~\eqref{eq:block-packing} with $b=(1,\ldots,1)$ was studied by
Grigoriadis and Khachiyan~\cite{GrigoriadisKhachiyan1994}.
For a constant block-solver tolerance, their randomized algorithm uses
$O(k\log m(\varepsilon^{-2}+\log k))$ expected block-solver calls.
Their deterministic algorithm has the same number of outer
iterations, but every iteration makes $k$ parallel block-solver calls.
Leighton et al.~\cite{LeightonEtAl1995} instantiate the same principle
for concurrent multicommodity flow by randomly selecting a commodity, whereas
Radzik~\cite{Radzik1997} replaces random selection by a fixed
round-robin order in that setting.  Villavicencio and
Grigoriadis~\cite{VillavicencioGrigoriadis1996}
establish cyclic
derandomization for this general block model with nonnegative
coupling constraints.  Thus, when $b=(1,\ldots,1)$, they obtain the same
asymptotic deterministic cyclic bound as Theorem~\ref{thm:main}; their stated bound is
$O(k\log m(\varepsilon^{-2}+\log\min\{k,m\}))$ block-solver calls.\footnote{
The models in~\cite{GrigoriadisKhachiyan1994,VillavicencioGrigoriadis1996}
are slightly different from ours: they
optimize over truncated blocks and impose different width conditions.
For C-MCF, these distinctions do not affect the resulting guarantees.
}
This common-sign structure permits a monotone
potential argument.  The present analysis retains the black-box block
oracle and cyclic order while allowing both packing and covering rows.

\myparagraph{General packing and covering}
More generally, multiplicative-weights and Lagrangian-relaxation methods have long been
used for packing and covering problems; Arora, Hazan, and
Kale~\cite{AroraHazanKale2012} give a broad survey.  In particular, for
$k=1$, the mixed-sign guarantee follows from their standard MWU framework
for convex feasibility with an asymmetric bounded oracle; see
\cite[Theorems~3.3-3.5]{AroraHazanKale2012}.
Alternative width-independent bounds for $k=1$, under different global
oracle specifications and for more general convex packing and concave
covering functions, were given by Jansen~\cite{Jansen2006} and Diedrich and
Jansen~\cite{DiedrichJansen2007}.
Plotkin, Shmoys, and
Tardos~\cite{PlotkinShmoysTardos1995} developed an oracle-based framework
for fractional packing and covering.  Young~\cite{Young2001} and
Quanrud~\cite{Quanrud2020} studied MWU algorithms for explicitly
represented mixed packing and covering LPs, without the block-oracle
structure assumed here.  In particular, Young gave sequential and
parallel width-independent algorithms, while Quanrud obtained a
deterministic algorithm with nearly linear running time.
For pure packing and covering, Klein and Young~\cite{KleinYoung2015}
proved an $\Omega(\rho\varepsilon^{-2}\log m)$ lower bound for
Dantzig-Wolfe-type algorithms in the applicable parameter regime.  Thus,
for $k=1$, the dependence of Theorem~\ref{thm:main} on
$\rho,\varepsilon$, and $m$ is optimal for that class of algorithms; their
result does not address the dependence on $k$ or the restriction to cyclic
individual-block updates.

\subsection{Bit complexity}
To achieve the bit-complexity claim in Theorem~\ref{thm:main}, we store each
internal weight as a floating-point number with logarithmic-bit mantissa and
exponent, rounding it multiplicatively after every update.  This accommodates
the potentially exponential dynamic range of the weights.  Prices passed to
the oracle are instead ordinary rationals with logarithmic-bit numerators and
a common denominator, obtained by truncation before normalization.  We call
this a \emph{hybrid weight-price representation}.  We are not aware of
earlier MWU work that explicitly uses this representation.

For MWU variants with exponential updates, an alternative would be to maintain
rounded rational logarithms of the weights.  The update then consists of
rational addition followed by rounding to a fixed grid; exponentiation is
needed only when computing the normalized prices.  Hedge~\cite{FreundSchapire1997} is a basic example of this
exponential-update viewpoint.  Implementations of the exponentiation step,
including softmax and
matrix-exponential computation, have been studied in
\cite{BlanchardHighamHigham2021,Kale2007Thesis,
TsudaRatschWarmuth2005,OrecchiaSachdevaVishnoi2012,
MuscoMuscoSidford2018,CarmonDuchiSidfordTian2019}.  In general, these works
do not ask whether both the maintained state and the normalized prices
supplied to a black-box oracle can be represented by rationals of logarithmic
bit length with a short common denominator.

Grigoriadis and Khachiyan~\cite[Section~9.1]{GrigoriadisKhachiyan1994}
form prices by normalizing exponentials of the current resource loads and
explicitly address the price side of this question.  They state that their
general block algorithm retains its iteration bound when the normalized price
components are computed to logarithmic precision; the remaining bit analysis
depends on the block solver.  Their minimum-cost-flow application rounds
prices to a rational grid of logarithmic bit
length~\cite{GrigoriadisKhachiyan1996MinCost}.

An exponentiation-free special case is Weighted Majority with $\beta=1/2$: its weights are
powers of two encoded by logarithmic-bit mistake counts and updated by binary
shifts~\cite{LittlestoneWarmuth1994}.  It has no optimization oracle, so it
does not address the representation of oracle prices.

\section{Algorithm and guarantees}

For a sequence of selected blocks $(i_t)_{t=1}^T$, write
\[
 \mathcal T_i:=\{t\in[T]:i_t=i\}
\]

\begin{algorithm}[H]
  \DontPrintSemicolon
  \SetNoFillComment
  set $w^{(1)}={\bf 1}\in\mathbb R^m$ \\
  \For{$t=1,\ldots,T$}
  {
    choose $i_t\in[k]$ \\
    $x^{(t)}\leftarrow {\tt Oracle}(i_t,y^{(t)})$ with $y^{(t)}=w^{(t)}/\sum_e w_e^{(t)}$ \\
    set $r^{(t)}=A_{i_t}x^{(t)}-b/k$ \\
    set $w_e^{(t+1)}=w_e^{(t)}(1+\eta r_e^{(t)})$
  }
  return $x=(x_1,\ldots,x_k)$ where
  $x_i=|\mathcal T_i|^{-1}\sum_{t\in\mathcal T_i}x^{(t)}$
  \caption{MWU for block-separable feasibility.}
  \label{alg:block-mwu}
\end{algorithm}

MWU-R chooses every $i_t$ independently and uniformly from $[k]$.
MWU-D uses the deterministic rule
$i_t=1+((t-1)\operatorname{mod}k)$.

\begin{theorem}\label{thm:mwu-r}
Suppose that \eqref{eq:block-packing} is feasible,
$\varepsilon\in(0,1/2]$, $\delta\in(0,1)$,
$\eta=\frac{\varepsilon}{10\rho}$, and
$
 T\ge48k\rho\varepsilon^{-2}
 \left(\ln\frac{2m}{\delta}+\ln\frac{2k}{\delta}\right)
$.
Then MWU-R outputs a $3\varepsilon$-approximate solution with probability at least $1-\delta$.
\end{theorem}

\begin{theorem}\label{thm:mwu-d}
Suppose that \eqref{eq:block-packing} is feasible, $\varepsilon\in(0,1/2]$, $\eta=\frac{\varepsilon}{10\rho}$, and $T\ge24k\rho\varepsilon^{-2}\ln(2m)$
is a multiple of $k$.
Then MWU-D outputs a $4\varepsilon$-approximate solution.
\end{theorem}

\myparagraph{Bit complexity}
Vectors $w$ in Algorithm~\ref{alg:block-mwu} may grow to rational numbers requiring polynomially many bits to store.
Next, we discuss how this can be avoided.

For integers $A,B\ge 1$, define
\[
 \mathbb Q_{A,B}
 :=\left\{
   \frac p{2^B}\cdot 2^{a}:
   p,a\in\mathbb Z,\ |p|<2^B,\ |a|<2^A
  \right\}
\]
Thus, numbers in $\mathbb Q_{A,B}$ can be represented with $O(A+B)$ bits ($O(A)$ bits for the exponent and $O(B)$ bits for the mantissa).
For value $z\in\mathbb R$ let ${\tt round}_B(z)$ be a number in $\bigcup_{A\ge1}\mathbb Q_{A,B}$ of the form $z(1+\delta)$ with $|\delta|\le 2^{-B}$; clearly, such number always exists.
Finally, define ${\tt truncate}_\tau(z)=\begin{cases}z\mbox{ if }|z|\ge \tau \\ 0 \mbox{ if }|z|<\tau\end{cases}$.
We extend these operations to vectors by applying them componentwise.

\begin{theorem}\label{thm:mwu-d-floating}
Suppose that \eqref{eq:block-packing} is feasible,
$\varepsilon\in(0,1/6]$, $\eta=3\varepsilon/(10\rho)$, and
$T\ge 3k\rho\varepsilon^{-2}\ln(2m)$ is a multiple of $k$.  Use
the block oracle with additive accuracy $\varepsilon/k$.
Consider the following modification of MWU-D: (1) compute $y^{(t)}$ at line 4 as follows: 
set $W^{(t)}=\sum_e w_e^{(t)}$, $\bar w^{(t)}={\tt truncate}_{\tau W^{(t)}}(w^{(t)})$, and $y^{(t)}=\bar w^{(t)}/\sum_e\bar w_e^{(t)}$; 
(2) after computing $w_e^{(t+1)}$ at line 6, update $w_e^{(t+1)}:={\tt round}_B(w_e^{(t+1)})$.

If
$
 B=\Theta(\log(k\rho T/\varepsilon))
$ and
$ \tau=\Theta(\varepsilon/(mk\rho))
$
with appropriate constants, then MWU-D produces a
$12\varepsilon$-approximate solution after
$T=\Theta(k\rho\varepsilon^{-2}\log m)$ oracle calls.  During the
algorithm all weights $w_e^{(t)}$ belong to $\mathbb Q_{A,B}$ for
$A=\Theta(\log(T\eta\rho+1))$, while every oracle vector has the form
$y^{(t)}=z^{(t)}/q_t$, where $z^{(t)}\in\mathbb Z_{\ge0}^m$,
$\sum_ez_e^{(t)}=q_t$, and
$q_t\le(mk\rho T/\varepsilon)^{O(1)}$.
\end{theorem}

Note that Theorem~\ref{thm:main} follows from the theorems above by running
the algorithms for the specified number of iterations, checking the returned
point, and reporting infeasibility if the check fails.

\section{Proofs of Theorems~\ref{thm:mwu-r},~\ref{thm:mwu-d},
and~\ref{thm:mwu-d-floating}}

By convexity of sets $\calX_i$, the output of Algorithm~\ref{alg:block-mwu} satisfies $x\in \calX_1\times\ldots\times \calX_k$.

For $t\ge1$ let us define
$$
 \Phi_t:=\sum_{e\in[m]}w_e^{(t)}
 \qquad
 \Phi_t^+:=\sum_{e\in E^+}w_e^{(t)}
 \qquad
 \Phi_t^-:=\sum_{e\in E^-}w_e^{(t)}
$$
The proofs will follow the usual strategy used for MWU: first, we prove that potential $\Phi_t$ does not grow too fast with $t$,
and then for each $e\in[m]$ we will use condition $w_e^{(T+1)}\le\Phi_{T+1}$ to conclude that $(Ax)_e\le b_e + O(\varepsilon)$.

The oracle specification implies that
\begin{equation}\label{eq:alternative-residual-ranges}
 \begin{alignedat}{2}
 r_e^{(t)}&\in[-1/k,\rho-1/k] &\qquad &\forall e\in E^+\\
 r_e^{(t)}&\in[-\rho+1/k,1/k] &&\forall e\in E^-
 \end{alignedat}
\end{equation}
Consequently,
\begin{equation}\label{eq:alternative-residual-squares}
 \begin{alignedat}{2}
 (r_e^{(t)})^2&\le\rho\bigl(r_e^{(t)}+2/k\bigr)
     &\qquad &\forall e\in E^+\\
 (r_e^{(t)})^2&\le\rho\bigl(-r_e^{(t)}+2/k\bigr)
     &&\forall e\in E^-
 \end{alignedat}
\end{equation}

\subsection{Analysis of MWU-R}

\myparagraph{Upper-bounding $\Phi_{T+1}$}
Let $x^*$ be a feasible solution.  Approximate optimality implies
\begin{equation}\label{eq:alternative-r-oracle}
 \sum_i y^\top A_i \cdot \mathtt{Oracle}(i,y)
 \le \sum_i y^\top A_i x_i^*+\varepsilon
 \le y^\top b+\varepsilon
\end{equation}

Let $\mathcal F_{t-1}$ be the history before $i_t$ is sampled and set
$S^{(t)}=\langle y^{(t)},r^{(t)}\rangle$.  By
\eqref{eq:alternative-r-oracle},
\[
 \mathbb E[S^{(t)}\mid\mathcal F_{t-1}]
 =\frac1k\left(
       \sum_i{y^{(t)}}^\top A_i\cdot\mathtt{Oracle}(i,y^{(t)})
       -{y^{(t)}}^\top b\right)
 \le\frac{\varepsilon}{k}
\]
Because $\Phi_{t+1}=\Phi_t(1+\eta S^{(t)})$, it follows that
\[
 \mathbb E[\Phi_{t+1}\mid\mathcal F_{t-1}]
 \le(1+\eta\varepsilon/k)\Phi_t
\]
Iterating from $\Phi_1=m$ gives
\[
 \mathbb E[\Phi_{T+1}]
 \le m(1+\eta\varepsilon/k)^T
 \le m\exp\left(\frac{\eta\varepsilon T}{k}\right)
\]
Markov's inequality therefore gives
\begin{equation}\label{eq:alternative-r-upper}
 \Phi_{T+1}\le \frac{2m}{\delta}
       \exp\left(\frac{\eta\varepsilon T}{k}\right)
\end{equation}
with probability at least $1-\delta/2$.

\myparagraph{Proof of Theorem~\ref{thm:mwu-r}}
Fix $e\in[m]$, and define $R_e=\sum_{t=1}^T r_e^{(t)}$.  
Suppose that $R_e\ge 0$. 
Using eq.~\eqref{eq:alternative-residual-squares} and bound
$\ln(1+\eta r)\ge\eta r-\eta^2r^2$, we get
\begin{equation}\label{eq:alternative-r-lower}
 \ln\Phi_{T+1}
 \ge\ln w_e^{(T+1)}
 \ge\frac{3\eta}{4}R_e-\frac{2\eta^2\rho T}{k}
\end{equation}
Combining \eqref{eq:alternative-r-upper} and
\eqref{eq:alternative-r-lower} and then using definitions of $\eta$ and $T$ yields
\[
 \frac{k R_e}{T}
 \le \frac 83\eta\rho
 +\frac43\frac{k\ln(2m/\delta)}{\eta T}
 +\frac43 \varepsilon
 \le \frac{17}{9}\varepsilon=:s
\]
If $R_e<0$, then $kR_e/T\le s$ holds automatically.  Thus
$kR_e/T\le s$ for every $e\in[m]$.
Chernoff's inequality and a union bound show, with
probability at least $1-\delta/2$, that
\begin{equation}\label{eq:GLADKJGHAK}
 |\mathcal T_i|=(1\pm\varepsilon/4)T/k
 \qquad\forall i\in[k]
\end{equation}
Assume that \eqref{eq:alternative-r-upper} and \eqref{eq:GLADKJGHAK} are both true; this holds with probability at least $1-\delta$.
By the definition of $r^{(t)}$ and $R_e$,
\[
 \sum_{t=1}^T(A_{i_t}x^{(t)})_e
 =R_e+\frac{Tb_e}{k}
 =\begin{cases}
   R_e+T/k,&e\in E^+\\
   R_e-T/k,&e\in E^-
  \end{cases}
\]
For $e\in E^+$ we have $(A_i x^{(t)})_e\ge 0$ and
$1/|\mathcal T_i|\le k/((1-\varepsilon/4)T)$. Therefore,
\[
 \begin{aligned}
 \sum_i(A_i x_i)_e
 =\sum_i\frac1{|\mathcal T_i|}
      \underbrace{\sum_{t\in\mathcal T_i}(A_i x^{(t)})_e}_{\ge 0}
 \le\frac{k}{(1-\varepsilon/4)T}
      (R_e+T/k)
 =\frac{1+kR_e/T}{1-\varepsilon/4}
 \le\frac{1+s}{1-\varepsilon/4}
 \le1+3\varepsilon
 \end{aligned}
\]
Similarly, for $e\in E^-$ we have $(A_i x^{(t)})_e\le0$ and
$1/|\mathcal T_i|\ge k/((1+\varepsilon/4)T)$.  Therefore,
\[
 \begin{aligned}
 \sum_i(A_i x_i)_e
 =\sum_i\frac1{|\mathcal T_i|}
      \underbrace{\sum_{t\in\mathcal T_i}(A_i x^{(t)})_e}_{\le 0}
 \le\frac{k}{(1+\varepsilon/4)T}
      (R_e-T/k)
 =\frac{-1+kR_e/T}{1+\varepsilon/4}
 \le\frac{-1+s}{1+\varepsilon/4}
 \le-1+3\varepsilon
 \end{aligned}
\]

\subsection{Analysis of MWU-D}

\myparagraph{Upper-bounding $\Phi_{T+1}$} We start with the following result.

\begin{lemma}\label{lem:cyclic-comparator}
There exists a vector $q=(q_1,\ldots,q_k)$ such that
\begin{subequations}
\begin{align}
 \sum_iq_i&=0
 \label{eq:cyclic-comparator-zero-sum}\\
 q_{i,e}&\ge-(1+\varepsilon)/k
 &&\forall i\in[k],\ e\in E^+
 \label{eq:cyclic-comparator-positive}\\
 q_{i,e}&\le1/k
 &&\forall i\in[k],\ e\in E^-
 \label{eq:cyclic-comparator-negative}\\
 \langle w^{(t)},r^{(t)}\rangle
 &\le\langle w^{(t)},q_i\rangle
      +\frac{2\varepsilon}{k}\Phi_t
 &&\forall t\in\mathcal T_i
 \label{eq:alternative-d-comparator}
\end{align}
\end{subequations}
\end{lemma}

\begin{proof}
For each block, define
\[
 \mathcal R_i=\overline{\operatorname{conv}}
 \{A_i\cdot{\tt Oracle}(i,y):y\in\Delta_m\}
\]
By the width conditions, $\mathcal R_i$ is contained in a bounded box;
since it is closed by definition, it is compact.  Hence
$\mathcal R_i+\mathbb R_+^m$ is a closed convex upper set.
For every $v\in A_i\calX_i$, there is a $z\in\mathcal R_i$ satisfying
$z\le v+(\varepsilon/k){\bf1}$.  Indeed, otherwise strict separation of
$v+(\varepsilon/k){\bf1}$ from the closed convex upper set
$\mathcal R_i+\mathbb R_+^m$ would give a separating vector
$y\ge0$, since the set is upper closed.  Normalizing $y$ gives
$y\in\Delta_m$ and
\[
 y^\top v+\varepsilon/k<\min_{z\in\mathcal R_i}y^\top z
\]
contradicting approximate oracle optimality.
Also, for every $y\in\Delta_m$ and $z\in\mathcal R_i$,
$\min_{x\in\mathcal X_i}y^\top A_ix\le y^\top z$: this holds for the
convex hull because $\mathcal X_i$ is convex and extends to its closure
by continuity.

Fix a feasible tuple $x_i^*$ and choose $z_i\in\mathcal R_i$ with
$z_i\le A_ix_i^*+(\varepsilon/k){\bf1}$.  Initially set
\[
 q_i=z_i-b/k-(\varepsilon/k){\bf1}
\]
Feasibility of $x^*$ and the choice of the $z_i$ give
\[
 \sum_iq_i
 =\sum_i z_i-b-\varepsilon{\bf1}
 \le\sum_iA_ix_i^*-b
 \le0
\]
Moreover, every vector in $\mathcal R_i$ satisfies $z_{i,e}\ge0$ on
$E^+$ and $z_{i,e}\le0$ on $E^-$.  Thus the initial vectors satisfy
$q_{i,e}\ge-(1+\varepsilon)/k$ on $E^+$
and 
$q_{i,e}\le(1-\varepsilon)/k$ on $E^-$.  Increase coordinates of the $q_i$
until \eqref{eq:cyclic-comparator-zero-sum} holds.  On $E^+$ this
preserves \eqref{eq:cyclic-comparator-positive}.  On $E^-$ it can be
done while preserving \eqref{eq:cyclic-comparator-negative}: the total available increase
$\sum_i(1/k-q_{i,e})=1-\sum_iq_{i,e}$ is at least the required increase
$-\sum_iq_{i,e}$.

It remains to show~\eqref{eq:alternative-d-comparator}.  First consider
the initial vectors $q_i$, before any coordinates are increased.  For
$t\in\mathcal T_i$, since $z_i\in\mathcal R_i$, we have
$
 \min_{x\in\mathcal X_i}\langle y^{(t)},A_ix\rangle
 \le \langle y^{(t)},z_i\rangle
$.
Approximate optimality of 
$x^{(t)}={\tt Oracle}(i,y^{(t)})$ therefore yields
\[
 \langle y^{(t)},A_ix^{(t)}\rangle
 \le \min_{x\in\mathcal X_i}\langle y^{(t)},A_ix\rangle
       +\frac{\varepsilon}{k}
 \le \langle y^{(t)},z_i\rangle+\frac{\varepsilon}{k}
\]
Since $y^{(t)}=w^{(t)}/\Phi_t$, multiplying by $\Phi_t$ gives
\[
 \langle w^{(t)},A_ix^{(t)}\rangle
 \le \langle w^{(t)},z_i\rangle+\frac{\varepsilon}{k}\Phi_t
\]
Using $r^{(t)}=A_ix^{(t)}-b/k$ and the initial definition
$q_i=z_i-b/k-(\varepsilon/k){\bf1}$, we therefore obtain
\[
 \langle w^{(t)},r^{(t)}\rangle
 \le \langle w^{(t)},q_i\rangle+\frac{2\varepsilon}{k}\Phi_t
\]
which is \eqref{eq:alternative-d-comparator}.  Increasing $q_i$
preserves this condition because $w^{(t)}\ge0$.
\end{proof}

Fix a vector $q$ given by Lemma~\ref{lem:cyclic-comparator}.
Since $\sum_{i=1}^k q_{i,e}=0$ for each $e\in [m]$, we conclude that there exists a $k$-periodic sequence $\{Q_{t,e}\}_{t=0,1,\ldots}$
satisfying
$
 q_{i_t,e}=Q_{t,e}-Q_{t-1,e}
$.
By adding a constant to $Q_{t,e}$ we can assume that $\max_{t}Q_{t,e}=0$ for $e\in E^+$ and $\min_{t}Q_{t,e}=0$ for $e\in E^-$.
Conditions~\eqref{eq:cyclic-comparator-positive} and~\eqref{eq:cyclic-comparator-negative}
now imply that  $Q_{t,e}\in[-(1+\varepsilon),0]$ on $E^+$ and
$Q_{t,e}\in[0,1]$ on $E^-$.
We can write
\[
 \sum_{t=1}^Tw_e^{(t)}q_{i_t,e}
 =w_e^{(T)}Q_{T,e}-w_e^{(1)}Q_{0,e}
 +\sum_{t=1}^{T-1}(w_e^{(t)}-w_e^{(t+1)})Q_{t,e}
\]
The update rule implies
$
 w_e^{(t)}-w_e^{(t+1)}=-\eta w_e^{(t)}r_e^{(t)}
$, and hence 
\[
 \sum_{t=1}^Tw_e^{(t)}q_{i_t,e}
 =w_e^{(T)}Q_{T,e}-w_e^{(1)}Q_{0,e}
 -\eta\sum_{t=1}^{T-1}w_e^{(t)}r_e^{(t)}Q_{t,e}
\]
Conditions $Q_{t,e}\in[-(1+\varepsilon),0]$ for $e\in E^+$ and
$Q_{t,e}\in[0,1]$ for $e\in E^-$ now imply
\begin{subequations}\label{eq:cyclic-coordinate-bounds}
\begin{align}
 \sum_{t=1}^Tw_e^{(t)}q_{i_t,e}
	&\le
        \makebox[2.5em][c]{$\displaystyle 0$}
        +
        \makebox[5.5em][c]{$\displaystyle (1+\varepsilon)w_e^{(1)}$}
        +
        \makebox[13em][c]{$\displaystyle
          \eta(1+\varepsilon)\sum_{t=1}^{T}w_e^{(t)}(r_e^{(t)})_+$}
	&& \forall e\in E^+
       \\
 \sum_{t=1}^T w_e^{(t)}q_{i_t,e}
       &\le
        \makebox[2.5em][c]{$\displaystyle w_e^{(T)}$}
        +
        \makebox[5.5em][c]{$\displaystyle 0$}
        +
        \makebox[13em][c]{$\displaystyle
          \eta\sum_{t=1}^{T}w_e^{(t)}(-r_e^{(t)})_+$}
	&& \forall e\in E^-
\end{align}
\end{subequations}
Here we extended both sums from $T-1$ to $T$, which is valid because
the added summands are nonnegative.
We have $w_e^{(T+1)}=w_e^{(T)}(1+\eta r_e^{(T)})$ and $r_e^{(T)}\ge-\rho$, and hence
$
 w_e^{(T)}
 \le \frac{w_e^{(T+1)}}{1-\eta\rho}
$.
Define
\[
\begin{array}{ll}
P_T^+=\displaystyle\sum_{t=1}^T\sum_{e\in E^+}w_e^{(t)}(r_e^{(t)})_+,&
N_T^+=\displaystyle\sum_{t=1}^T\sum_{e\in E^+}w_e^{(t)}(-r_e^{(t)})_+\\
P_T^-=\displaystyle\sum_{t=1}^T\sum_{e\in E^-}w_e^{(t)}(r_e^{(t)})_+,&
N_T^-=\displaystyle\sum_{t=1}^T\sum_{e\in E^-}w_e^{(t)}(-r_e^{(t)})_+
\end{array}
\]
Thus $P_T^\pm$ and $N_T^\pm$ collect the positive and negative parts,
respectively, of the residual contributions.  Summing the two bounds
in \eqref{eq:cyclic-coordinate-bounds} over their respective coordinate
sets gives
\[
 \sum_{t=1}^T\langle w^{(t)},q_{i_t}\rangle
 \le (1+\varepsilon)\Phi_1^+
      +\frac{\Phi_{T+1}^-}{1-\eta\rho}
      +\eta\bigl((1+\varepsilon)P_T^++N_T^-\bigr)
\]
Finally, summing \eqref{eq:alternative-d-comparator} over
$t=1,\ldots,T$ and substituting the preceding bound yields
\begin{equation}\label{eq:alternative-d-charge}
 \sum_{t=1}^T\langle w^{(t)},r^{(t)}\rangle
 \le(1+\varepsilon)\Phi_1^++\frac{\Phi_{T+1}^-}{1-\eta\rho}
 +\eta\bigl((1+\varepsilon)P_T^++N_T^-\bigr)
 +\frac{2\varepsilon}{k}\sum_{t=1}^T\Phi_t
\end{equation}

The update rule also gives
\begin{align*}
 \Phi_{T+1}-\Phi_1
 &=\eta\sum_{t=1}^T\langle w^{(t)},r^{(t)}\rangle\\
 \Phi_{T+1}^+-\Phi_1^+&=\eta(P_T^+-N_T^+)\\
 \Phi_{T+1}^--\Phi_1^-&=\eta(P_T^--N_T^-)
\end{align*}
The last two identities give
$
 P_T^+=N_T^++\frac{\Phi_{T+1}^+-\Phi_1^+}{\eta}
$ 
and
$
 N_T^-=P_T^-+\frac{\Phi_1^--\Phi_{T+1}^-}{\eta}
$.  Substituting these identities into \eqref{eq:alternative-d-charge}
and collecting the $\Phi_{T+1}^+$ and $\Phi_{T+1}^-$ terms yields
\begin{align}
 &(1-\eta(1+\varepsilon))\Phi_{T+1}^+
 +\left(1-\frac{\eta^2\rho}{1-\eta\rho}\right)\Phi_{T+1}^-\nonumber \\
 &\qquad\le \Phi_1^++(1+\eta)\Phi_1^-
 +\eta^2\bigl((1+\varepsilon)N_T^++P_T^-\bigr)
 +\frac{2\eta\varepsilon}{k}\sum_{t=1}^T\Phi_t
 \label{eq:NGLASHFLAS}
\end{align}
We can write
\begin{equation}\label{eq:GNLASJHFAIG}
1-\eta(1+\varepsilon)
\le 1-\frac{\eta^2\rho}{1-\eta\rho}
\qquad\quad
 (1+\varepsilon)N_T^++P_T^-
 \le\frac{1+\varepsilon}{k}\sum_{t=1}^T\Phi_t
\qquad\quad
\Phi_1=m
\end{equation}
(The first bound holds since $\eta\rho=\varepsilon/10\le1/20$ and hence
$1+\varepsilon\ge \frac{\eta\rho}{1-\eta\rho}$).
Dividing~\eqref{eq:NGLASHFLAS} by $1-\eta(1+\varepsilon)$ and using~\eqref{eq:GNLASJHFAIG} gives
\[
 \Phi_{T+1}\le \alpha m+\beta\sum_{t=1}^T\Phi_t
 \qquad\mbox{where}
 \quad
 \alpha:=\frac{1+\eta}{1-\eta(1+\varepsilon)}
 \qquad
 \beta:=
 \frac{\eta^2(1+\varepsilon)+2\eta\varepsilon}
 {k[1-\eta(1+\varepsilon)]}
\]
Iterating this recurrence starting from $\Phi_1=m$ gives
\begin{equation}\label{eq:alternative-d-upper}
 \Phi_{T+1}\le \alpha m(1+\beta)^T
 \le \alpha m e^{\beta T}
\end{equation}
where the first inequality is proved by induction on $T$.

\myparagraph{Proof of Theorem~\ref{thm:mwu-d}}
Fix $e\in[m]$, and again define $R_e=\sum_{t=1}^T r_e^{(t)}$.  Suppose that $R_e\ge0$.
As in eq.~\eqref{eq:alternative-r-lower}, we get 
\begin{equation}\label{eq:alternative-d-lower}
 \ln\Phi_{T+1}
 \ge\ln w_e^{(T+1)}
 \ge\frac{3\eta}{4}R_e-\frac{2\eta^2\rho T}{k}
\end{equation}
Comparing \eqref{eq:alternative-d-upper} with
\eqref{eq:alternative-d-lower} gives
\begin{equation}\label{eq:alternative-d-residual-preliminary}
 \frac{R_e}{T}
 \le \frac83\frac{\eta\rho}{k}
 +\frac{4\beta}{3\eta}
 +\frac{4\ln(\alpha m)}{3\eta T}
\end{equation}

Recall that $\eta\le\eta\rho=\varepsilon/10$ and
$\varepsilon\le1/2$.  We first claim that
$$
 \frac{\beta k}{\eta}
 = \frac{\eta(1+\varepsilon)+2\varepsilon}
 {1-\eta(1+\varepsilon)}
 <\frac{7\varepsilon}{3}
$$
This holds since the denominator is positive and the difference of cross products is negative:
$$
 3\bigl(\eta(1+\varepsilon)+2\varepsilon\bigr)
 -7\varepsilon\bigl(1-\eta(1+\varepsilon)\bigr)
 =\eta(1+\varepsilon)(3+7\varepsilon)-\varepsilon
 \le\frac{\varepsilon}{10}\cdot\frac{39}{4}
             -\varepsilon
 <0
$$
We also have
$\eta\le\varepsilon/10\le1/20$, 
$\eta(1+\varepsilon)\le3/40$ and $T\ge 24k\rho\varepsilon^{-2}\ln(2m)=\frac{24k}{10\eta\varepsilon}\ln(2m)$, and hence
\[
 \alpha
 =\frac{1+\eta}{1-\eta(1+\varepsilon)}
 \le\frac{21/20}{37/40}
 <2
\qquad\quad
 \frac{\ln(\alpha m)}{\eta T}
 \le\frac{\ln(2m)}{\eta T}
 \le\frac 5{12}\frac{\varepsilon}{k}
\]
Substituting these bounds 
into \eqref{eq:alternative-d-residual-preliminary} gives
\begin{equation}\label{eq:alternative-d-residual}
 \frac{R_e}{T}
 \le 
 \;
 \frac 83 \cdot \frac{\varepsilon}{10k} 
 \; + \;
 \frac 43 \cdot \frac{7\varepsilon}{3k} 
 \; + \;
 \frac 43 \cdot \frac{5\varepsilon}{12k} 
 \;
 < \frac{4\varepsilon}{k}
\end{equation}
We established bound~\eqref{eq:alternative-d-residual} assuming that $R_e\ge 0$. If $R_e<0$
then the same bound holds automatically. Thus,~\eqref{eq:alternative-d-residual} holds for all $e\in[m]$.
This finally implies
\[
 \sum_iA_i x_i-b
 =\frac{k}{T}\sum_{t=1}^T r^{(t)}
 \le4\varepsilon{\bf1}
\]

\subsection{Finite-precision implementation: proof of Theorem~\ref{thm:mwu-d-floating}}

Let $\widetilde w^{(t)}$ and $\widetilde y^{(t)}$ 
be the weights and the oracle vectors in the modified algorithm,
and $x^{(t)}$ be the oracle answer in the $t$-th step.
Let $w^{(t)}$ and $y^{(t)}$ be similar quantities 
in the execution of the original MWU-D algorithm with parameter $3\varepsilon$ in which the oracle returns 
$x^{(t)}$ in the $t$-th step. It suffices to show that this answer fits the specification
of the approximate oracle, i.e.\ that
\begin{equation}\label{eq:GALIUHGKAJSHF}
(y^{(t)})^\top A_i x^{(t)} \le \min_{x\in\calX_i}(y^{(t)})^\top A_i x+3\varepsilon/k
\end{equation}
for each $t\in[T]$ and $i=i_t$. Theorem~\ref{thm:mwu-d} will then imply that the output is an $4\cdot(3\varepsilon)$-approximate solution.

Since the two executions use the same
oracle answer $x^{(t)}$, they also use the same residual
$
 r^{(t)}=A_i x^{(t)}-b/k
$.
By construction, we have $w_e^{(1)}=\widetilde w_e^{(1)}=1$ and
\begin{equation*}
 w_e^{(t+1)}=w_e^{(t)}(1+\eta r_e^{(t)})
 \qquad\qquad
 \widetilde w_e^{(t+1)}=\widetilde w_e^{(t)}(1+\eta r_e^{(t)})(1+\delta_{t,e})
\end{equation*}
for some $\delta_{t,e}$ with $|\delta_{t,e}|\le \delta:=2^{-B}$.
This implies that 
\begin{equation}\label{eq:GALHKHFALSG}
 \left|\log_2\frac{\widetilde w_e^{(t)}}{w_e^{(t)}}\right|
 \le T\max\{ |\log_2 (1+\delta)|, |\log_2 (1-\delta)| \}
 \le 2T\delta
\end{equation}
for each $t\in[T]$.  Next, we compare $y:=y^{(t)}$ and $\widetilde y:=\widetilde y^{(t)}$. Recall that they are constructed
from $w:=w^{(t)}$ and $\widetilde w:=\widetilde w^{(t)}$ as follows:
\[
 y=\frac{w}{\sum_e w_e}
 \qquad
\widetilde W=\sum_e\widetilde w_e
 \qquad
\overline w={\tt truncate}_{\tau\widetilde W}
(\widetilde w)
 \qquad
\overline W=\sum_e\overline w_e
\qquad
 \widetilde y
 =\frac{\overline w}{\overline W}
\]
We choose the constant in $\tau=\Theta(\varepsilon/(mk\rho))$ so that
$\tau<1/m$.  Since some coordinate of $\widetilde w$ is at least
$\widetilde W/m$, at least one coordinate survives the truncation;
hence $\overline W>0$ and $\widetilde y$ is well defined.
The definitions of $y$ and $\widetilde y$ will not change if vectors $w$ and $\tilde w$ are divided by the same constant;
this operation will also preserve~\eqref{eq:GALHKHFALSG}. We can thus assume w.l.o.g.\ that $\sum_e w_e=1$.
This condition together with \eqref{eq:GALHKHFALSG} implies
$\widetilde w_e \in w_e\cdot [2^{-2T\delta},2^{2T\delta}]$
and hence 
$\|\widetilde w-w\|_1\le 2^{2T\delta}-1$.
Choose the constant in $B=\Theta(\log(k\rho T/\varepsilon))$ so that
$2^{2T\delta}-1\le\varepsilon/(8k\rho)$; then
$|\widetilde W-1|\le
\|\widetilde w-w\|_1\le\varepsilon/(8k\rho)$.
Choose the constant in $\tau=\Theta(\varepsilon/(mk\rho))$ so that
$
 \|\overline w-\widetilde w\|_1
 =\widetilde W-\overline W
 \le m\tau\widetilde W
 \le\frac{\varepsilon}{8k\rho}
$.
We obtain
$$
|\overline W-1| \le \|\overline w-w\|_1 \le \|\overline w-\widetilde w\|_1 + \|\widetilde w-w\|_1
\le \frac{\varepsilon}{8k\rho} + \frac{\varepsilon}{8k\rho} = \frac{\varepsilon}{4k\rho}
$$

Since $y=w$ and $\widetilde y=\overline w/\overline W$ under our
normalization, we obtain 
\begin{equation}\label{eq:finite-precision-stability}
\begin{aligned}
 \|\widetilde y-y\|_1
 \le
 \left\|\frac{\overline w}{\overline W}-\overline w\right\|_1
 +\|\overline w-w\|_1
 =|1-\overline W|+\|\overline w-w\|_1
 \le\frac{\varepsilon}{2k\rho}
\end{aligned}
\end{equation}

Let $x_i^\star={\tt Oracle}_{\varepsilon}(i,y)$ where the subscript indicates the oracle's accuracy.
(Recall that $x:=x^{(t)}={\tt Oracle}_{\varepsilon}(i,\widetilde y)$).
  The width bounds and
\eqref{eq:finite-precision-stability} imply
\[
 \left|(\widetilde y-y)^\top A_i z\right|
 \le\frac{\varepsilon}{2k}
 \qquad\text{for }z\in\{x,x_i^\star\}
\]
Using this estimate and the oracle guarantee at
$\widetilde y$ and $y$, we obtain the desired inequality~\eqref{eq:GALIUHGKAJSHF}:
\[
\begin{aligned}
 y^\top A_i x
 &\le \widetilde y^\top A_i x
       +\frac{\varepsilon}{2k}\\
 &\le \widetilde y^\top A_i x_i^\star
       +\frac{3\varepsilon}{2k}\\
 &\le y^\top A_i x_i^\star
       +\frac{2\varepsilon}{k}\\
 &\le \min_{x\in\calX_i}y^\top A_i x
       +\frac{3\varepsilon}{k}
\end{aligned}
\]

It remains to discuss the ranges of $\widetilde w^{(t)}$ and
$\widetilde y^{(t)}$ in the modified algorithm.  First,
$|\log_2\widetilde w_e^{(t)}|=O(T(\eta\rho+\delta))$, so the exponent
of an internal weight has $O(\log(T\eta\rho+1))$ bits.  Together with
the $B$-bit mantissa produced by rounding, this gives the stated
membership in $\mathbb Q_{A,B}$.

We next consider $\widetilde y^{(t)}$.  Any two nonzero coordinates of
$\overline w^{(t)}$ differ by a factor of at most
$1/\tau=O(mk\rho/\varepsilon)$: each is at least
$\tau\widetilde W$, while each is at most $\widetilde W$.  Write every
nonzero rounded weight in canonical dyadic form
$p_e2^{a_e-B}$ with $2^{B-1}\le p_e<2^B$ (absorbing powers of two into
$a_e$).  The preceding ratio bound implies
$|a_e-a_{e'}|=O(\log(1/\tau))$ for nonzero retained coordinates.  After
factoring out $2^{\min_e a_e-B}$, the retained weights are integers
$z_e=p_e2^{a_e-\min_ea_e}$ having
$O(B+\log(1/\tau))$ bits.  Therefore
\[
 \widetilde y_e=\frac{z_e}{q},\qquad
 q=\sum_ez_e,
\]
where both $z_e$ and $q$ have
$O(B+\log m+\log(1/\tau))=
O(\log(mk\rho T/\varepsilon))$ bits.  This gives the stated
polynomially bounded shared denominator.

\section{Maximum weighted multicommodity flow}
In this section we apply Theorem~\ref{thm:main} to prove Theorem~\ref{thm:m-mcf}.
Let $G=(V,E)$ be a directed or undirected graph with capacities $u_e>0$.
Commodity $i$ has terminals $(s_i,t_i)$ and reward $\gamma_i>0$ per unit
of flow.
Edges of zero capacity and commodities of zero reward can be discarded,
so these positivity assumptions are without loss of generality.
For a candidate objective value $F>0$, let $\calX_i$ consist of pairs
$x_i=(f_i,\beta_i)$ such that $f_i$ is an $s_i$-$t_i$ flow of value
$\beta_i$, $0\le f_i(e)\le u_e$ for every $e\in E$, and
$0\le\gamma_i\beta_i\le F$.  
The feasibility problem for value $F$ can be written as follows:
\begin{subequations}\label{eq:mcf-feasibility}
\begin{align}
 x=(x_1,\ldots,x_k) \in & \; \calX_1\times\ldots\times \calX_k \\
 \sum_{i=1}^k \frac{f_i(e)}{u_e}&\le 1
       &&\forall e\in E\label{eq:mcf-capacity}\\
 -\sum_{i=1}^k\frac{\gamma_i\beta_i} F&\le -1\label{eq:mcf-reward}
\end{align}
\end{subequations}
This target system is feasible if and only if $F\le\mathrm{OPT}$.
The definition of $\calX_i$ implies that any vector $x_i\in\calX_i$ satisfies
conditions \eqref{eq:approx-oracle:b}, \eqref{eq:approx-oracle:c} with width parameter $\rho=1$.  Given prices $y$, the $i$-th oracle
needs to solve
\begin{equation}\label{eq:C-objective}
\min_{(f,\beta)\in\calX_i} C(f,\beta),\qquad C(f,\beta)
 :=y_\circ +\sum_{e\in E}\frac{y_e}{u_e}f_e
 -\frac{y_\circ\gamma_i}{F}\beta
\end{equation}
where $y_\circ$ is the component corresponding to constraint~\eqref{eq:mcf-reward}.
(We shifted the objective by $y_\circ$, which does not affect the oracle and its additive approximation guarantees).
Note that $C(f,\beta)\ge 0$ for any $(f,\beta)\in\calX_i$ since $y_\circ\left(1-\frac{\gamma_i}{F}\beta\right)\ge 0$.

We can model~\eqref{eq:C-objective} as a min-cost flow problem as follows.
Define $D=F/\gamma_i$.
Add a parallel dummy edge $(s_i,t_i)$ (directed or undirected, depending on the type of the multicommodity flow problem),
 and seek a
minimum-cost $s_i$-$t_i$ flow of value exactly $D$ in the resulting
graph.  Give every original edge $e$ capacity $u_e$ and cost
$y_e/u_e$, and give the dummy edge capacity $D$ and cost
$y_\circ/D$.  If $\beta$ units use the original graph, the remaining
$D-\beta$ units use the dummy edge, and the resulting cost is
exactly $C(f,\beta)$.  Thus all costs in this
min-cost-flow instance are nonnegative.  In the undirected case, cancel
cycles in the returned flow before reading off $(f,\beta)$.  In
particular, any backward use of the dummy edge forms, together with an
$s_i$-$t_i$ path in the original graph, a nonnegative-cost cycle and
can be removed without increasing cost.  We may therefore assume that
the dummy edge carries an amount in $[0,D]$ in the forward direction,
so the remaining original flow has value $\beta\in[0,D]$ as required.

Sending all $D$ units through the dummy edge has cost $y_\circ$,
so $C^\ast:=\min_{(f,\beta)\in\calX_i} C(f,\beta)\le y_\circ\le 1$.  Hence,
if the min-cost-flow solver finds feasible flow $(f,\beta)$ within factor $1+\varepsilon/k$ from the optimum
then its additive error is $\varepsilon/k$, as needed:
\[
 C(f,\beta)
 \le(1+\varepsilon/k)C^\ast
 \le C^\ast+\varepsilon/k
\]

We showed that a feasible target $F$ can be tested with $O(k\varepsilon^{-2}\log m)$
min-cost-flow calls; if problem~\eqref{eq:mcf-feasibility} is feasible then we find an $O(\varepsilon)$-approximate solution.
Scaling all flows uniformly makes the capacity constraints exact while
retaining total reward of the flow at least $(1-O(\varepsilon))F$.

This is a gap test: an infeasibility report implies $F>\mathrm{OPT}$,
whereas a returned solution can be scaled to a feasible flow of value
$(1-O(\varepsilon))F$.  Treating these outcomes as the upper and lower
answers, respectively, gives a multiplicative gap search.
If $\mathrm{OPT}=0$, the zero flow is already optimal.  Otherwise,
polynomial boundedness implies that the optimum is at least
$(mk)^{-O(1)}$, while the elementary bound
$\mathrm{OPT}\le\sum_i\gamma_i\sum_eu_e$ is at most
$(mk)^{O(1)}$.
Searching to relative accuracy $1+O(\varepsilon)$ uses
$O(\log(km/\varepsilon))$ tests.  This establishes
Theorem~\ref{thm:m-mcf}.


\begin{remark}
In the introduction we claimed an efficient algorithm for families with efficient vertex separators
via the work~\cite{DongEtAl2022}.
Note that adding the dummy edge does not cause a problem for such families: if $S$ is a vertex separator of size $O(\sqrt n)$ in the original
graph, then $S\cup\{s_i\}$ is a separator of the augmented graph (also of size $O(\sqrt n)$). 
\end{remark}

\subsection*{Acknowledgements and declaration on the use of generative AI}
During the preparation and revision of this manuscript, the author used ChatGPT and Codex by OpenAI as interactive research and writing tools. 
They were used to generate draft proof arguments, including initial drafts of the proofs of Theorems~\ref{thm:mwu-r}, \ref{thm:mwu-d}, 
and~\ref{thm:mwu-d-floating}, and to assist with literature searches, consistency checks, language editing, and presentation.

All research questions, algorithmic framework, and final mathematical arguments were developed and selected by the author. 
All AI-generated material was independently checked and substantially revised by the author; references and prior-work claims
were verified against the cited sources. The author takes full responsibility for the correctness and content of the manuscript.

This work was in part motivated by interactions with Jack Spalding-Jamieson, during which the author learned
about the existence of a (less efficient) reduction of M-MCF to min-cost flows.

\bibliographystyle{plain}
\bibliography{MWU}

\begin{thebibliography}{10}

\bibitem{AroraHazanKale2012}
Sanjeev Arora, Elad Hazan, and Satyen Kale.
\newblock The multiplicative weights update method: A meta-algorithm and
  applications.
\newblock {\em Theory of Computing}, 8(6):121--164, 2012.
\newblock \url{https://doi.org/10.4086/toc.2012.v008a006}.

\bibitem{BlanchardHighamHigham2021}
Pierre Blanchard, Desmond~J. Higham, and Nicholas~J. Higham.
\newblock Accurate computation of the log-sum-exp and softmax functions.
\newblock {\em IMA Journal of Numerical Analysis}, 41(4):2311--2330, 2021.
\newblock \url{https://doi.org/10.1093/imanum/draa038}.

\bibitem{CarmonDuchiSidfordTian2019}
Yair Carmon, John~C. Duchi, Aaron Sidford, and Kevin Tian.
\newblock A rank-1 sketch for matrix multiplicative weights.
\newblock In {\em Proceedings of the 32nd Conference on Learning Theory},
  volume~99 of {\em Proceedings of Machine Learning Research}, pages 589--623,
  2019.
\newblock \url{https://proceedings.mlr.press/v99/carmon19a.html}.

\bibitem{ChenGraurSidford2025}
Li~Chen, Andrei Graur, and Aaron Sidford.
\newblock Accelerated approximate optimization of multi-commodity flows on
  directed graphs.
\newblock In {\em Proceedings of the 57th Annual ACM Symposium on Theory of
  Computing}, pages 2203--2212, 2025.
\newblock \url{https://doi.org/10.1145/3717823.3718267}.

\bibitem{ChenEtAl2025MinCostFlow}
Li~Chen, Rasmus Kyng, Yang~P. Liu, Richard Peng, Maximilian Probst~Gutenberg,
  and Sushant Sachdeva.
\newblock Maximum flow and minimum-cost flow in almost-linear time.
\newblock {\em Journal of the ACM}, 72(3):19:1--19:103, 2025.
\newblock Conference version at FOCS 2022.
  \url{https://doi.org/10.1145/3728631}.

\bibitem{DiedrichJansen2007}
Florian Diedrich and Klaus Jansen.
\newblock An approximation algorithm for the general mixed packing and covering
  problem.
\newblock In {\em Combinatorics, Algorithms, Probabilistic and Experimental
  Methodologies}, volume 4614 of {\em Lecture Notes in Computer Science}, pages
  128--139, 2007.
\newblock \url{https://doi.org/10.1007/978-3-540-74450-4_12}.

\bibitem{DongEtAl2022}
Sally Dong, Yu~Gao, Gramoz Goranci, Yin~Tat Lee, Richard Peng, Sushant
  Sachdeva, and Guanghao Ye.
\newblock Nested dissection meets {IPM}s: Planar min-cost flow in nearly-linear
  time.
\newblock In {\em Proceedings of the 2022 Annual ACM-SIAM Symposium on Discrete
  Algorithms}, pages 124--153, 2022.
\newblock \url{https://doi.org/10.1137/1.9781611977073.7}.

\bibitem{Fleischer2000}
Lisa~K. Fleischer.
\newblock Approximating fractional multicommodity flow independent of the
  number of commodities.
\newblock {\em SIAM Journal on Discrete Mathematics}, 13(4):505--520, 2000.
\newblock Conference version at FOCS 1999.
  \url{https://doi.org/10.1137/S0895480199355754}.

\bibitem{FreundSchapire1997}
Yoav Freund and Robert~E. Schapire.
\newblock A decision-theoretic generalization of on-line learning and an
  application to boosting.
\newblock {\em Journal of Computer and System Sciences}, 55(1):119--139, 1997.
\newblock Conference version at EuroCOLT 1995.
  \url{https://doi.org/10.1006/jcss.1997.1504}.

\bibitem{GargKonemann1998}
Naveen Garg and Jochen K{\"o}nemann.
\newblock Faster and simpler algorithms for multicommodity flow and other
  fractional packing problems.
\newblock In {\em Proceedings of the 39th Annual IEEE Symposium on Foundations
  of Computer Science}, pages 300--309, 1998.
\newblock \url{https://doi.org/10.1109/SFCS.1998.743463}.

\bibitem{GrigoriadisKhachiyan1994}
Michael~D. Grigoriadis and Leonid~G. Khachiyan.
\newblock Fast approximation schemes for convex programs with many blocks and
  coupling constraints.
\newblock {\em SIAM Journal on Optimization}, 4(1):86--107, 1994.
\newblock \url{https://doi.org/10.1137/0804004}.

\bibitem{GrigoriadisKhachiyan1995Exponential}
Michael~D. Grigoriadis and Leonid~G. Khachiyan.
\newblock An exponential-function reduction method for block-angular convex
  programs.
\newblock {\em Networks}, 26(2):59--68, 1995.
\newblock \url{https://doi.org/10.1002/net.3230260202}.

\bibitem{GrigoriadisKhachiyan1996MinCost}
Michael~D. Grigoriadis and Leonid~G. Khachiyan.
\newblock Approximate minimum-cost multicommodity flows in {$\widetilde
  O(\varepsilon^{-2}KNM)$} time.
\newblock {\em Mathematical Programming}, 75(3):477--482, 1996.
\newblock Technical report published in 1995.
  \url{https://doi.org/10.1007/BF02592195}.

\bibitem{GrigoriadisKhachiyan1996Coordination}
Michael~D. Grigoriadis and Leonid~G. Khachiyan.
\newblock Coordination complexity of parallel price-directive decomposition.
\newblock {\em Mathematics of Operations Research}, 21(2):321--340, 1996.
\newblock \url{https://doi.org/10.1287/moor.21.2.321}.

\bibitem{Jansen2006}
Klaus Jansen.
\newblock Approximation algorithm for the mixed fractional packing and covering
  problem.
\newblock {\em SIAM Journal on Optimization}, 17(2):331--352, 2006.
\newblock Conference version at IFIP TCS 2004.
  \url{https://doi.org/10.1137/030601570}.

\bibitem{Kale2007Thesis}
Satyen Kale.
\newblock {\em Efficient Algorithms Using the Multiplicative Weights Update
  Method}.
\newblock PhD thesis, Princeton University, 2007.
\newblock Technical Report TR-804-07.
  \url{https://www.satyenkale.com/papers/thesis.pdf}.

\bibitem{Karakostas2008}
George Karakostas.
\newblock Faster approximation schemes for fractional multicommodity flow
  problems.
\newblock {\em ACM Transactions on Algorithms}, 4(1):13:1--13:17, 2008.
\newblock Conference version at SODA 2002.
  \url{https://doi.org/10.1145/1328911.1328924}.

\bibitem{KleinYoung2015}
Philip Klein and Neal~E. Young.
\newblock On the number of iterations for {Dantzig--Wolfe} optimization and
  packing-covering approximation algorithms.
\newblock {\em SIAM Journal on Computing}, 44(4):1154--1172, 2015.
\newblock Conference version at IPCO 1999.
  \url{https://doi.org/10.1137/12087222X}.

\bibitem{KleinPlotkinSteinTardos1994}
Philip~N. Klein, Serge~A. Plotkin, Clifford Stein, and {\'E}va Tardos.
\newblock Faster approximation algorithms for the unit capacity concurrent flow
  problem with applications to routing and finding sparse cuts.
\newblock {\em SIAM Journal on Computing}, 23(3):466--487, 1994.
\newblock Conference version at STOC 1990.
  \url{https://doi.org/10.1137/S0097539792241175}.

\bibitem{LeightonEtAl1995}
Tom Leighton, Fillia Makedon, Serge Plotkin, Clifford Stein, {\'E}va Tardos,
  and Spyros Tragoudas.
\newblock Fast approximation algorithms for multicommodity flow problems.
\newblock {\em Journal of Computer and System Sciences}, 50(2):228--243, 1995.
\newblock Conference version at STOC 1991.
  \url{https://doi.org/10.1006/jcss.1995.1020}.

\bibitem{LittlestoneWarmuth1994}
Nick Littlestone and Manfred~K. Warmuth.
\newblock The weighted majority algorithm.
\newblock {\em Information and Computation}, 108(2):212--261, 1994.
\newblock Conference version at FOCS 1989.
  \url{https://doi.org/10.1006/inco.1994.1009}.

\bibitem{Madry2010}
Aleksander Madry.
\newblock Faster approximation schemes for fractional multicommodity flow
  problems via dynamic graph algorithms.
\newblock In {\em Proceedings of the 42nd ACM Symposium on Theory of
  Computing}, pages 121--130, 2010.
\newblock \url{https://doi.org/10.1145/1806689.1806708}.

\bibitem{MuscoMuscoSidford2018}
Cameron Musco, Christopher Musco, and Aaron Sidford.
\newblock Stability of the {Lanczos} method for matrix function approximation.
\newblock In {\em Proceedings of the 29th Annual ACM--SIAM Symposium on
  Discrete Algorithms}, pages 1605--1624, 2018.
\newblock \url{https://doi.org/10.1137/1.9781611975031.105}.

\bibitem{OrecchiaSachdevaVishnoi2012}
Lorenzo Orecchia, Sushant Sachdeva, and Nisheeth~K. Vishnoi.
\newblock Approximating the exponential, the {Lanczos} method and an
  {$\widetilde O(m)$}-time spectral algorithm for balanced separator.
\newblock In {\em Proceedings of the 44th Annual ACM Symposium on Theory of
  Computing}, pages 1141--1160, 2012.
\newblock \url{https://doi.org/10.1145/2213977.2214080}.

\bibitem{PlotkinShmoysTardos1995}
Serge~A. Plotkin, David~B. Shmoys, and {\'E}va Tardos.
\newblock Fast approximation algorithms for fractional packing and covering
  problems.
\newblock {\em Mathematics of Operations Research}, 20(2):257--301, 1995.
\newblock Conference version at FOCS 1991.
  \url{https://doi.org/10.1287/moor.20.2.257}.

\bibitem{Quanrud2020}
Kent Quanrud.
\newblock Nearly linear time approximations for mixed packing and covering
  problems without data structures or randomization.
\newblock In {\em Proceedings of the 2020 ACM--SIAM Symposium on Simplicity in
  Algorithms}, pages 69--80, 2020.
\newblock \url{https://doi.org/10.1137/1.9781611976014.11}.

\bibitem{Radzik1997}
Tomasz Radzik.
\newblock Fast deterministic approximation for the multicommodity flow problem.
\newblock {\em Mathematical Programming}, 78(1):43--58, 1997.
\newblock Conference version at SODA 1995.
  \url{https://doi.org/10.1007/BF02614505}.

\bibitem{ShahrokhiMatula1990}
Farhad Shahrokhi and David~W. Matula.
\newblock The maximum concurrent flow problem.
\newblock {\em Journal of the ACM}, 37(2):318--334, 1990.
\newblock Conference version at the ACM Computer Science Conference 1987.

\bibitem{Sherman2017}
Jonah Sherman.
\newblock Area-convexity, $\ell_\infty$ regularization, and undirected
  multicommodity flow.
\newblock In {\em Proceedings of the 49th Annual ACM Symposium on Theory of
  Computing}, pages 452--460, 2017.
\newblock \url{https://doi.org/10.1145/3055399.3055501}.

\bibitem{TsudaRatschWarmuth2005}
Koji Tsuda, Gunnar R\"atsch, and Manfred~K. Warmuth.
\newblock Matrix exponentiated gradient updates for on-line learning and
  {Bregman} projection.
\newblock {\em Journal of Machine Learning Research}, 6:995--1018, 2005.
\newblock \url{https://www.jmlr.org/papers/v6/tsuda05a.html}.

\bibitem{VillavicencioGrigoriadis1996}
Jorge Villavicencio and Michael~D. Grigoriadis.
\newblock Approximate structured optimization by cyclic block-coordinate
  descent.
\newblock In Herbert Fischer, Bruno Riedm{\"u}ller, and Stefan Sch{\"a}ffler,
  editors, {\em Applied Mathematics and Parallel Computing: Festschrift for
  Klaus Ritter}, pages 359--371. Physica-Verlag, Heidelberg, 1996.
\newblock \url{https://doi.org/10.1007/978-3-642-99789-1}.

\bibitem{Young2001}
Neal~E. Young.
\newblock Sequential and parallel algorithms for mixed packing and covering.
\newblock In {\em Proceedings of the 42nd IEEE Symposium on Foundations of
  Computer Science}, pages 538--546, 2001.
\newblock \url{https://doi.org/10.1109/SFCS.2001.959930}.

\end{thebibliography}

\end{document}